# Speech-Based Emotion Recognition using Neural Networks and Information Visualization

Jumana Almahmoud*  
MIT, CSAIL

Kruthika Kikkeri**  
MIT, RLE and MTL

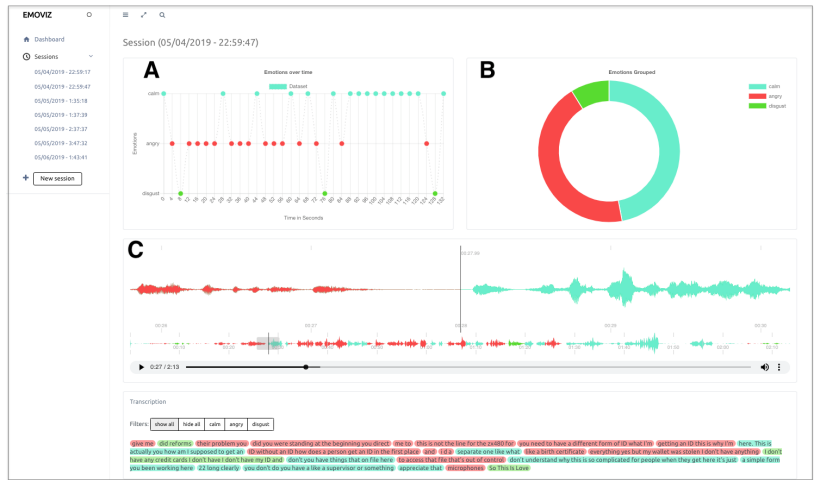

Figure 1: EmoViz Dashboard. A: visualization shows emotions over time during a session. B: shows the overall emotions in a certain session. C: The waveform is colored based on emotions and mapped to the text bellow that could be filtered based on emotions.

## ABSTRACT

Emotions recognition is commonly employed for health assessment. However, the typical metric for evaluation in therapy is based on patient-doctor appraisal. This process can fall into the issue of subjectivity, while also requiring healthcare professionals to deal with copious amounts of information. Thus, machine learning algorithms can be a useful tool for the classification of emotions. While several models have been developed in this domain, there is a lack of user-friendly representations of the emotion classification systems for therapy. We propose a tool which enables users to take speech samples and identify a range of emotions (happy, sad, angry, surprised, neutral, clam, disgust, and fear) from audio elements through a machine learning model. The dashboard is designed based on local therapists' needs for intuitive representations of speech data in order to gain insights and informative analyses of their sessions with their patients.

## 1  INTRODUCTION

Emotions play a key role in human life as they affect both human physiology and psychology. In therapeutic settings, improvement relies on how coherently patients express their emotions and how well the therapist recognizes and reacts to these utterances [2]. However, a therapist manages many patients for an extended period of time, which introduces a high volume of data that needs to be managed and maintained. Thus, a platform which can provide informative, speech-based emotion recognition insights can be useful for a variety of applications including therapy sessions. We propose EmoViz (Figure 1), a dashboard which enables users to take speech samples and identify a range of emotions (e.g. happy, sad, angry, surprised, neutral, etc.) through a neural network. Through this analysis of speech signals, emotion information can be detected without the use of invasive systems such as facial recognition or internal-signal based physiological readouts (i.e. EKG, ECG, MRI) [3,4,6]. The platform, EmoViz, provide users with a dashboard that visualizes informative analyses of the range of emotions over time, as well as clustering audio and texts based on emotions. EmoViz serves as a management system for therapists to track their sessions by taking advantage of speech recognition technology, emotional recognition and data visualization.

*  jumanam@mit.edu.  
** kkikkeri@mit.edu.

## 2  SYSTEM COMPONENTS

The EmoViz system consists of three modules: the background, the backend, and the frontend. The background processes the neural network models and the audio transcription by taking in audio files and creating segments of emotions to associated with the transcription. Figure 2 is an illustration of the system architecture. EmoViz is open source and available online at https://github.com/JumanaFM/emoviz.

### 2.1  Emotion Corpus

Audio files of actors expressing a range of emotions was employed as an emotion corpus [5]. This emotion corpus, known as the Ryerson Audio-Visual Database of Emotional Speech and Song (RAVDESS), consisted of 24 actors (12male, 12 female), who expressed eight different emotions: neutral, calm, happy, sad, angry, fearful, disgust, and sur-prised. Each actor articulated 60 unique clips for a total of 1440 audio files. The data was split into a training dataset of 1008 files and 432 test files for optimization of model. An additional emotion corpus, the Toronto Emotional Speech Set (TESS), derived from [1] was also utilized for further experimentation. In contrast to the original training and testing dataset, the TESS corpus was comprised of two actresses (age 26 and 64) who enacted seven different emotions: anger, disgust, fear, happiness, pleasant surprise, sadness, and neutral.

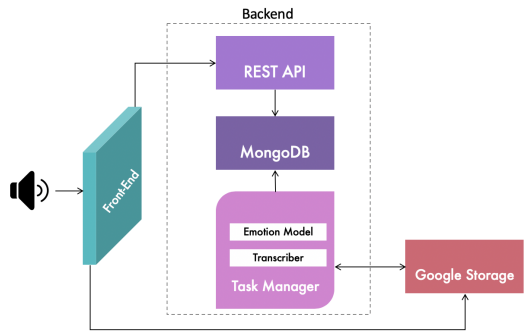

Figure 2: EmoViz System Architecture

## 2.2 Experimental Methods

To devise the emotion recognition model for this platform, we investigated three approaches for the neural network architecture: Deep Neural Network (DNN), with an input of audio features. Convolution Neural Network (CNN), with an input of audio features. DNN/CNN with audio and text input features. Table 1 summarizes the error rates. For DNN and CNN models without text analysis for epochs of 600. Note that the error was calculated as the total number of false detections divided by the total of detections for each emotion. The accuracy is the 1-error rate.

Table 1. Summary of emotion classification error rate

| Emotion | DNN (Err %) | CNN (Err %) |
| --- | --- | --- |
| Neutral | 23 | 25 |
| Calm | 12 | 22 |
| Happy | 4 | 3 |
| Sad | 9 | 5 |
| Angry | 6 | 10 |
| Fearful | 11 | 14 |
| Disgust | 15 | 17 |
| Surprised | 10 | 12 |

## 3 DASHBOARD

In order to support therapists' workflow, we built the EmoViz interface. EmoViz is a system for emotion analysis based on speech signals. In the following sections we describe the dashboard and the user experience.

### 3.1 Platform Description

EmoViz provide three different charts that provide insights about the emotional composition of audio files provided by the user. As shown in Figure 1, one visualization is a connection between the waveform and the transcribed text. The user can jump to specific emotions or filter out others. Users can also hover over a certain emotion on the waveform to display corresponding text from the recorded session. The other two visualizations display the overall emotions in the session and emotions over time.

For emotions over time, we display the x axis as time, and y axis as different emotions. For the donut chart, we calculated the sum of emotions that appeared during a session to give users a holistic view of the session. Finally, for the audio and transcribed text, we granted users control over this visualization, by giving them the ability to connect the audio with the text and navigate to the emotions concurrently. Additionally, we mapped the color of each emotion on the waveform and the text. This enables a user to hover over a waveform and see the corresponding text without having to scroll. The three panels of visualization are connected and could be cross filtered to represent a recorded session through different lenses (time, emotion, aggregate).

### 3.2 Usage Pipeline

Therapists can record a session using their preferred method of recording. They can then log into EmoViz and access the history of sessions for an individual patient. They can then add a new session for that patient by uploading the audio recording. The audio recording will then be internally processed such that the speech file will be transcribed into text and highlighted by the color of emotion for corresponding times and phrases. A summary of the emotions will also be displayed on the dashboard. The therapist can then filter through these emotions or explore them overtime. The user experience of EmoViz provides a richer way to reflect on a session through the chart panels by incorporating the transcribed text and recordings into the visualizaitons.

### 3.3 Evaluation with Therapists

We consulted two therapists at a local mental health centre. We performed an informal study with them and asked them about what their thoughts about the overall experience. We received positive feedback, as they explained that this system has a huge potential in the field of therapy to manage their patients. They also mention their interest in using the tool in a group therapy setting as quoted by one therapist: "Often it can be challenging to keep track of everyone's emotion during a group session. We try to look at everyone's face, but we can't keep track of everyone." The other therapist also mentioned: "I'm just thinking that this could be really useful in a group setting. You would just need everyone to consent but maybe it would be useful for the patients to have as well."

## 4 CONCLUSION

We explored the implementation of a neural network model for emotion detection. This was for the purposes of developing a user-friendly platform for therapy sessions. To execute this task, we explored a variety of different neural network topologies, and input features for emotion classification. The dashboard displays a summary of the emotions expressed over time and a highlighted audio signal and transcribed text. This enables users to listen to the audio as it highlights different emotions, as well read the transcripts of the signal. Further-more, this dashboard enables the user to identify and view single emotions at a time.

## REFERENCES

[1] Fabio Calefato, Filippo Lanubile, and Nicole Novielli. 2017. EmoTxt: A toolkit for emotion recognition from text. In *2017 Seventh International Conference on Affective Computing and Intelligent Interaction Workshops and Demos (ACIIW)*, IEEE, San Antonio, TX, 79–80. DOI:https://doi.org/10.1109/ACIIW.2017.8272591

[2] Christine L. Lisetti and Fatma Nasoz. 2004. Using Noninvasive Wearable Computers to Recognize Human Emotions from Physiological Signals. *EURASIP J. Adv. Sig. Proc.* 2004, (2004), 1672–1687. DOI:https://doi.org/10.1155/S1110865704406192

[3] Monorama Swain, Aurobinda Routray, and P. Kabisatpathy. 2018. Databases, features and classifiers for speech emotion recognition: a review. *Int J Speech Technol* 21, 1 (March 2018), 93–120. DOI:https://doi.org/10.1007/s10772-018-9491-z

[4] Noé Tits, Kevin El Haddad, and Thierry Dutoit. 2018. ASR-based Features for Emotion Recognition: A Transfer Learning Approach. In *Proceedings of Grand Challenge and Workshop on Human Multimodal Language (Challenge-HML)*, Association for Computational Linguistics, Melbourne, Australia, 48–52. Retrieved April 7, 2019 from https://www.aclweb.org/anthology/W18-3307

[5] T. Wehrle, S. Kaiser, S. Schmidt, and K. R. Scherer. 2000. Studying the dynamics of emotional expression using synthesized facial muscle movements. *J Pers Soc Psychol* 78, 1 (January 2000), 105–119.

[6] Lyssn: Intelligent Psychotherapy Assessment. Retrieved April 17, 2019 from https://lyssn.io/